\documentclass[nofootinbib]{revtex4-1}       

\usepackage{graphicx}
\usepackage{csquotes}
\usepackage[table]{xcolor}
\usepackage{amsmath}
\usepackage{amssymb}
\usepackage{epigraph}
\usepackage{physics}

\begin{document}

\title{Contingent free choice: on extensing quantum theory to a\\contextual, deterministic theory with improved predictive power}
\author{Ghislain Fourny}


\affiliation{ETH Zurich, Department of Computer Science}
\email{ghislain.fourny@inf.ethz.ch}

\date{Originally March 2019, Updated September 2019.}

\begin{abstract}
The non-extensibility of quantum theory into a theory with improved predictive power is based on a strong assumption of independent free choice, in which the physicists pick a measurement axis independently of anything that couldn't have been caused by their decision.

Independent free choice is also at the core of the Nash equilibrium and classical game theory. A more recent line of game-theoretical research based on weakening free choice leads to non-trivial solution concepts with desirable properties such as at-most uniqueness, Pareto optimality, and contextuality.

We show how introducing contingent free choice in the foundations of quantum theory yields a class of deterministic and contextual theories with an improved predictive power, and contrast them with the pilot-wave theory.

Specifically, we suggest that quantum experiments, such as the EPR experiment, involving measurements located in spacetime, can be recast as dynamic games with imperfect information involving human agents and the universe. The underlying idea is that a physicist picking a measurement axis and the universe picking a measurement outcome are two faces of the same physical contingency phenomenon.

The classical, Nashian resolution of these games based on independent free choice is analogous to local hidden variable theories, constrained by the Bell inequalities. On the other hand, in a setup in which agents are rational and omniscient in all possible worlds, under contingent free choice, the Perfectly Transparent Equilibrium provides a contextual resolution, based on the iterated elimination of inconsistent worlds, towards an at-most unique possible world, in which the outcomes of measurements that actually are carried out, and only them, are deterministically defined.

\end{abstract}

\maketitle

\section{Introduction}
\epigraph{If indeed there exist any experimenters with a modicum of free will, then elementary particles must have their own share of this valuable commodity.'}{\textit{John Conway, Simon Knochen (2006)}}

While there are numerous competing interpretation of quantum physics (Copenhagen, Many-Worlds, Quantum Bayesianism ...), most theoretical physicists are aligned on one fundamental assumption: free choice.

Under the assumption that an experimenter freely\footnote{We will make the meaning(s) of free choice precise in subsequent sections of this paper. In this sentence, we mean independent free choice, its stronger variant.} chooses a measurement axis, several contributions in the field of quantum physics have been made that demonstrate that quantum physics is inherently random, and thus cannot be extended to a theory with improved predictive power\footnote{The De Broglie-Bohm theory is the most prominent case of a deterministic theory; it does not fulfil the free choice assumption, like the candidate theory presented in this paper. However, the De-Broglie Bohm theory leaves the initial conditions open and does not explain why our world is the actual world, rather than another possible world with different initial conditions.}.

As \citet{Conway2006} put it, ``if indeed
there exist any experimenters with a modicum of free will, then elementary
particles must have their own share of this valuable commodity.'' 

However, and in spite of evidence in the field of neuroscience \citep{Libet1999} in contradiction with independent free choice, there was little research done so far on the consequences of weakening free choice in quantum theory, as opposed to completely dropping it.

In this note, we advocate that, beyond intimate convictions, independent free choice, which can be defined mathematically and formally, is an axiom that does not have to hold, and might possibly be experimentally proven not to hold. We advocate that a different approach in which a weaker version of free choice is assumed, contingent free choice, can potentially be equally consistent, and also useful.

We also contribute, to support this suggestion, an alternate deterministic, contextual and non-trivial theory with improved predictive power, based on the elimination of possible worlds that are not immune to perfect prediction under utility maximization, both by physicists carrying out experiments, and by the universe.\footnote{For the universe, quantities such as the quantum potential based on the Schrodinger equation in the De Broglie-Bohm theory, Fermat's principle in optics, etc, can be interpreted as utility maximization.}

\section{Newcomb's problem}

The notion of free choice, or free will, is probably as old as philosophy, and there are many ways that it can be defined. We argue in this paper that two equally reasonable definitions of free choice, independent free choice and contingent free choice, can be given and, importantly, that both of these definitions are formal and precise.

Free choice is often defined, or thought of, in contrast to the ability to predict what an agent endowed with free choice will decide before they do. This apparent incompatibility between being fully predictable and having free will is embodied in Newcomb's problem.

As has been argued in \citet{Gardner1973}, Newcomb's problem is to free will what Schrodinger's cat \citep{Schrodinger1935} is to quantum entanglement: it is a thought experiment that takes a seemlingly abstract notion and ties it to something tangible \footnote{Early mentions of these thoughts were made in a paper by Jon Lindsay written in J.-P. Dupuy's class at Stanford in 1994.}. Here a cat, there the content of a box.

\subsection{Formulation}

Newcomb's problem \footnote{It is also known as a paradox, however, we do not think it is one, as the two apparently conflicting resolutions, in the end, come down to the definition of free choice assumed: one box for contingent free choice, two boxes for independent free choice.} \citep{Gardner1973} is typically found under the following form.

An agent is presented with two boxes. One of the boxes is transparent, and it can be seen that it contains \$1,000. The other box is opaque and its contents cannot be seen, but it is known that it is either empty, or contains \$1,000,000.

The agent has the choice between either taking the opaque box, or both boxes. Whichever amounts are inside become hers. But there is one catch.

A while before this game took place, somebody predicted the agent's decision, and prepared the contents of the opaque box accordingly: if the predictor predicted that the agent would take one box, he put \$1,000,000 inside the opaque box. If, however, the predictor predicted that the agent would take two boxes, then he put nothing inside the opaque box. There is thus a well-defined causal dependency between the \emph{prediction} and the contents of the box.

Regarding the skills and accuracy of the prediction, past records of the game with other agents -- some of which took one box, some of which took two boxes -- are available, showing that the predictor has made one thousand correct predictions out of the one thousand games played so far.

What should the agent do?

\subsection{One box or two boxes?}

One line of reasoning, which can be qualified of Nashian \footnote{to refer to the work by John Nash}, is based on a dominant-strategy reasoning: calling $x$ the amount in the opaque box, utility is maximized by picking $1000+x$ over $x$, that is, both boxes should be taken. The people reasoning this way are casually named two-boxers. With a correct prediction, they get \$1,000.

Another line of reasoning which makes as much sense is that one box should be taken, leading to \$1,000,000 if the prediction is once again correct. This is because, if two boxes had been taken instead, then the prediction would have been correct too, that is, the predictor would have predicted that two boxes would be taken, and would have put nothing in the opaque box. Had the agent taken two boxes, he would have got only \$1,000 in total, which is less.

The resolution of this apparent paradox \citep{Dupuy1992}\citep{Fourny2018} lies in the modelling of the prediction, more exactly, in the counterfactuals: two-boxers assume that the prediction is correct, but would have been the same (and incorrect) if the agent had made the other choice. One-boxers assume that the prediction is correct and would also have been correct if the agent had made a different decision.

Both resolutions are thus as rational as each other, as agents maximize their utility under both reasonings, although the reasonings are based on different assumptions on the nature of free choice and the strength of the prediction.

\section{Counterfactuals}
\label{section-counterfactuals}

In the previous paragraph, the one-boxer line of reasoning was based on subjunctive conditionals: "if two boxes had been taken instead, then the prediction would have been correct too, that is, the predictor would have predicted that two boxes would be taken, and would have put nothing in the opaque box." This sort of implication is fundamental to the reasoning, and is called a counterfactual implication. It is paramount to understand that a counterfactual dependency is of a fundamentally different nature than a causal dependency. We now go into deeper details regarding counterfactual implications.

The one-boxer approach is often discarded on the grounds that only the first approach is consistent with causal consistency, because an event cannot cause another event that is not in its future light cone. However, as \citet{Dupuy1992} argues, causality is not the only kind of dependency: a dependency between two events may also be due to a correlation, to a quantum entanglement, or to any kind of counterfactual implication.

In the one-boxer approach, there is a counterfactual dependency between the decision and its prediction: I pick one box and it has been predicted I would pick one box, but if I had picked two boxes instead, it would have been predicted that I was going to pick two boxes.

\subsection{Counterfactual implications as opposed to logical implications}

Counterfactual implications have been formalized by \citet{Lewis1973} based on lining up alternate worlds around the actual world with the notion of a distance to the actual world. The counterfactual implication $A > B$ then means that, in the closest world where A is true, B is true as well. It is thus to be distinguished from a logical implication $A \implies B$ equivalent to $\neg A \vee  B$, which holds trivially in the actual world if A does not hold, but also from the notion of a necessary implication that would hold in all accessible\footnote{The notion of accessible world has been formally defined as part of the \citet{Kripke1963} semantics in the context of modal logics. Put simply, a world is epistemically accessible if it is compatible with an agent's knowledge in the actual world. It is logically accessible if it is compatible with the laws of logics applicable to the actual world. Depending on the definition taken, one can derive necessity operators ($\Box$) with different underlying semantics.} worlds: $\Box (A \implies B)$.

Formally, counterfactual implications are defined like so. There is a set of worlds $\Omega$. There is a notion of distance between these worlds. There are predicates on these worlds that may or may not hold. There is a function $f$ that associates each world $\omega$ and each predicate, for example $A$, to the closest world in which this predicate is true. The counterfactual implication $A>Q$ is said to hold if $Q$ is true in the world $f(\omega, A)$.

\subsubsection{Counterfactuals in Newcomb's problem}

Coming back to Newcomb's problem, the assertion by a one-boxer that ``if she had picked two boxes, the opaque box would be empty'' means that, in the closest world in which she picks two boxes, the predictor predicted so and put nothing inside the opaque box. 

For example, if the agent takes one box, the associated predicate that she takes one box, $A_1$, holds in the actual world $\omega$, which we write as follows:

$$\omega \models A_1$$

In this world, the prediction is that she will pick one box. Let us call this predicate (that the prediction is one box) $P_1$:

$$\omega \models A_1 \wedge P_1$$

If we call $A_2$ the predicate the agent picks two boxes, then the function $f$ maps $\omega$ and $A_2$ to some other world

$$\omega' = f(\omega, A_2)$$

and we have

$$\omega' \models A_2$$

by definition of f.

Note that we have $f(\omega, A_1)=\omega$, since $A_1$ already holds in $\omega$.

Then, if some predicate holds in $\omega'$, for example, that the prediction is made that the agent picked two boxes, $P_2$, then we can write:

$$\omega' \models P_2$$

or equivalently:

$$f(\omega, A_2) \models P_2$$

then the following counterfactual implication holds in $\omega$:

$$\omega \models A_2 > P_2$$

Expressed in other worlds \footnote{This is an obvious typo, but the author preferred to promote it to a pun rather than fix it.}, in the actual world $\omega$, it is true that the agent picks one box, that it was predicted she would pick one box, and that if she had picked two boxes, then it would have been predicated that she would pick two boxes, formally:

$$\omega \models A_1 \wedge P_1 \wedge (A_2 > P_2)$$

It can thus be seen that there is no causal implication directed to the past in this reasoning. A well known and broadly mentioned example of a non-causal, counterfactual dependency is when there exists a common cause. A counterfactual dependence that does not go into the future is called a backtracking counterfactual by \citet{Lewis1973}. Lewis actually defined causal dependency on top of the notion of counterfactual dependency (A causes B if it is true that, if A were not true, B would not be true either -- provided this is not a backtracking counterfactual). In this paper, we have the opposite approach of carefully distinguishing between the notion of causal dependency and that of counterfactual dependency. In this respect, for the purpose of this paper, A causes B if B is in A's future light cone\footnote{This is also the definition adopted by some theoretical physicists, e.g., \citet{Renner2011}.}, the precise relationship being modelled by an equation of motion such as Schrodinger's equation.

Likewise, the assertion by a two-boxer that ``if she had picked one box, the opaque box would be empty as well'' means that, in the closest world in which she picks one box, the predictor wrongly predicted she would pick two, and put nothing inside the opaque box.

Whether a counterfactual implication is true or not is thus a matter of assumption on the way possible worlds are organized. As we will see in Section \ref{section-free-choice}, it depends on whether we assume independent free choice (the past is the same in all possible worlds and my decision is independent of that past), or contingent free choice (the prediction is correct in all possible worlds, and I could have acted otherwise \footnote{Formally meaning: there (counterfactually) exists a world in which I act otherwise.}).

\subsubsection{Counterfactuals in quantum theory}

Another counterfactual dependency arises in quantum physics in the EPR experiment: the particles are prepared together to have them entangled, for example in the state

$$\ket{\psi}=\frac{\ket{\uparrow\uparrow}+\ket{\downarrow\downarrow}}{\sqrt{2}}$$

If two physicists get one particle and measure, space-like separated, along the same orthogonal basis $(\ket{\uparrow}, \ket{\downarrow})$, then they will find the same result, say, $\uparrow$. But quantum theory tells us that the following counterfactual implication materially holds: had one of the physicists measured $\downarrow$, then the other physicist would have measured $\downarrow$. Quantum theory is thus, by nature, a theory of counterfactuals \citep{Fuchs2013}. Formulated in the Lewisian framework: there exists a logically accessible world in which one the physicists measures $\downarrow$, and in the closest of these worlds, the other physicists also measures $\downarrow$.

\subsubsection{Counterfactuals as opposed to correlations}
\label{section-counterfactuals-correlations}

A solution to Newcomb's paradox based on a statistical framework, was also given by \citet{Baltag2009}. However, this solution is based on a Bayesian analysis and correlations rather than objective counterfactuals.

The difference between correlations and counterfactuals is at the core of a fundamental discussion in decision theory, between causal decision theory and evidential decision theory. In the former, agents maximize their expected utility with weights that are probabilities of subjunctive conditionals (counterfactuals), whereas in the latter, agents maximize their expected utility with weights that are conditional probabilities in the Bayesian sense \citep{Stalnaker1968}\citep{Stalnaker1972}.

Traditionally, causal decision theory is associated with the two-box solution and evidential decision theory with the one-box solution. However, it is important to point out that causal decision theory makes the independent free choice assumption (see Section \ref{section-free-choice}), in which anything not in the future light cone of a decision is counterfactually independent from it.

In this paper, our approach is that of causal decision theory in the sense that we consider counterfactuals, but with contingent free choice, which does not preclude backtracking counterfactual dependencies, i.e., going backwards in time or along spacelike-separated locations.

In the context of QBism \citep{Fuchs2013}, quantum theory is known to break the law of total probability, which is, given an event A and a partition of the Omega space $(B_i)_i$:

$$P(A) = \sum_i P(B_i) P(A|B_i)$$

As \citet{Fuchs2013} argues, this is also due to the counterfactual nature of quantum theory, as the events considered are, except one, counterfactual. If we explicitly rewrite $P(A|B_i)$ as a probability of subjunctive conditionals, then we immediately see that there is no reason why the following should always hold:

$$P(A) = \sum_i P(B_i) P(B_i > A)$$

The law of total probability is known not to hold for probabilities of subjunctive conditionals \citep{Fuchs2010}, with the Born rule being its quantum counterpart. Quantum Bayesianism is thus different from classical Bayesianism.

\section{Two definitions of free choice}
\label{section-free-choice}

Coming back to the Newcomb problem, the core divergence between the two reasonings is thus whether the prediction \emph{would have been} \footnote{The use of the conditional tense is paramount in counterfactual statements.} correct, counterfactually. It is precisely this which allows us to distinguish between two fundamental approaches to free choice.

\subsection{Independent free choice}

In the two-boxer approach, free choice means that the choice is uncorrelated to anything that does not lie in the future light cone of the decision \citep{Renner2011}\citep{Colbeck2017}. This is the strong version of free choice, which we call independent free choice.

This assumption is actually two-fold and involves that (i) the complement of the future light cone is counterfactually independent of the freely made decision and (ii) the complement to the future light cone is uncorrelated to the freely made decision.

More on the difference between counterfactual implications and correlations was formally given in Section \ref{section-counterfactuals-correlations}.

The first assumption is often only implicitly stated but is nevertheless just as fundamental for the second assumption to make sense: a setup with a decision uncorrelated to the complement of its future light cone in conditional probability terms, but in which the the complement of its future light is nevertheless counterfactually dependent on the decision would make very little sense and is certainly not the intent of the proponents of independent free choice. But let us make this argument more precise.

This would be theoretically possible if we defined the Lewis function $f$ in a way that it maps at least one world and one alternate decision to another world in which the complement of the future light cone is different. Indeed, $f$ is not the same mathematical object as the probability distribution on the possible worlds. However, Lewis pointed out that the notion of distance between two worlds should be tied to the volume of spacetime on which they fully coincide: the more they overlap, the closer they are. Thus, if B is statistically uncorrelated with A, then there must exist, by definition, a world for each possible combination of values of A and B on which the underlying probabilities for these single values are not zero. As a consequence, given any world, the closest world with a different value of A (or B) is one in which the other variable has the same value. This reasoning illustrates that statistical independence is tied with counterfactual independence in the context of (classical) causal decision theory.

\subsubsection{Counterfactual independence}

With this definition of free choice, the past is fixed and (in modal logics terms) necessary, so that any prediction made in the past can be made incorrect by the predicted agent simply by making a decision different than the prediction. In simple terms assuming non-relativistic time: there is nothing the agent can do at $t_2$ so that the prediction of his decision would have been different at $t_1<t_2$ \citep{Dupuy1992}. More formally, if we consider a possible world $w$ with a decision made at a given location in Minkowski spacetime, and if we call $Q$ any predicate on the complement of the future light cone, and $A_i$ the event that the choice made is $i$, then the following holds:

$$\forall i, Q \iff A_i > Q$$

Where $>$ is the symbol for counterfactual implication in the Lewisian sense (if the choice had been $A_i$, Q would be true). The above formula states that saying that "if choice $i$ were made, then Q would be true" is the same thing as saying that Q is true: the choice of $i$ has no counterfactual influence on the complement of the past light cone of the location of where this choice is made.

\subsubsection{No statistical correlation}

In setups involving probabilities, there may be more than one possible world that is consistent with the knowledge an agent has. A probability distribution exists on the set $\Omega$ of these possible worlds to model this uncertainty. In \citet{Kripke1963} semantics, these possible worlds and are also referred to as (epistemically) accessible worlds.

Independent free choice in this context involves the absence of any correlation between a decision event $A_i\subset \Omega$ and any predicate $Q$ (also an event, i.e., $Q\subset \Omega$) on the complement of the future light cone:

$$P(A_i|Q)=P(A_i)$$

or equivalently

$$P(Q|A_i)=P(Q)$$

If we now bring the counterfactual independence into this discussion, then it follows from

$$\forall i, Q \iff A_i > Q$$

that \footnote{Note that a counterfactual implication is also an event that may or may not be true in a given world, and so it can be assigned a probability.}:

$$\forall i, P(Q) = P(A_i > Q)$$

Thus, if the choice of $i$ is done freely and independently, then for any predicate $Q$ on the complement of the future light cone and possible decision $A_i$, we have:

$$\forall i, P(Q|A_i)=P(Q)=P(A_i > Q)$$

so that the difference between counterfactual independence and lack of correlation vanishes away. This explains why much of the literature based on independent free choice conflates the two kinds of dependencies in their definition of free choice and only writes that $P(Q|A_i)=P(Q)$ or the equivalent $P(A_i|Q)=P(A_i)$, implicitly meaning that the lack of correlation also implies counterfactual independence.

It is this approach to free choice that is commonly assumed in both Nashian game theory \citep{Nash1951}, where agents can consider unilateral modifications of their (pure or mixed) strategies, and in quantum physics \citep{Renner2011}, where physicists unpredictably pick their measurement axes.

\subsection{Contingent free choice}

In the one-boxer approach, free choice means that the agent could have acted otherwise, i.e., it means that for each considered choice $i$ the agent can make, there exists some world in which the agent makes choice $i$ \footnote{The terminology "full support" is also used to mean this in litterature \citep{Renner2011}.}. There is no restriction on any counterfactual dependency or correlation with other events not in the future light cone -- including in particular the prediction of their decision. In particular, if the agent makes decision $i$ in the actual world $\omega$, there is at least one another world $\omega'=f(\omega, j)$ in which he makes some other decision $j \neq i$, and the complement to the future light cone may look completely different in $\omega'$  than it does in $\omega$.

This is a weaker, compatibilistic version of free choice, which we call contingent free choice. It can be seen clearly from the careful use of terminology, that there is no causal relationship between $i$ or $j$ and the complement of the future light cone in $\omega$ or $\omega'$. This is a purely counterfactual relationship compatible with the laws of special relativity, with no signals travelling faster than light.

Assuming contingent free choice instead of independent free choice gives room for adding \footnote{but we do not have to.} the additional assumption that the prediction is correct in all possible worlds. This implies in particular that the prediction of a decision, say, a few hours earlier at the same location, or at the same time in a different room, is counterfactually dependent on that decision:

$$\forall i, A_i > \text{Prediction}(A_i)$$

In an ideal setup, which is the one considered in this paper as a simple illustration, the agents are perfect predictors, i.e., are epistemically omniscient in all possible worlds, i.e., the above applies to any event considered. As a consequence, there are no probabilities involved and we thus do not talk about correlations, but only about counterfactual dependencies\footnote{One \emph{could} view perfect prediction as a correlation between a decision and its prediction by considering the set of all possible worlds -- the actual one, and all counterfactual ones -- as some probability space endowed with an arbitrary probability distribution: the event of a decision and the event of its prediction perfectly match, which qualifies as a perfect correlation. However, we do not do so to avoid introducing more confusion than necessary.}.
 
In the actual world, the prediction is correct. In another, hypothetical world, the decision is different but the prediction is different as well, and it is also correct. In modal logics terms \citep{Kripke1963}, the prediction is \emph{necessarily} correct, but the past (or more generally the complement of the future light cone) is not \emph{necessary} as there are some worlds in which it is different: the past \emph{could have been} different.

\section{Proofs of non-extensibility of quantum theory}

Several proofs of the non-extensibility of quantum theory, or put equivalently, on the inherent randomness in quantum theory, are found in literature. These proofs all have in common a fundamental assumption: that the physicist performing a quantum measurement freely chooses the measurement axis, with the understanding of independent free choice. This underlying notion of free choice is based on a strong assumption that a decision taken freely is independent from the past. 

\subsection{The EPR experiment}

The EPR experiment serves as the basis for Bell's original theorem in this respect. EPR stands for Einstein, Podolsky and Rosen.

In the EPR experiment, two entangled particles are prepared and sent far away to two spacelike-separated physicists, Peter and Mary, who can make measurements, for example, on the polarization or on the spin. The initial state of the joint system is:

$$\ket{\phi}_{AB} = \frac{\ket{\uparrow\uparrow}_{AB}+\ket{\downarrow\downarrow}_{AB}}{\sqrt{2}}$$

The first physicist, Peter, measures the spin of his particle against an axis of his choice, say $(\ket{\uparrow}_A, \ket{\downarrow}_A)$. With 50\% of probability, he obtains $\uparrow$ and his half of the system collapses, from his perspective, to:

$$\ket{\phi'}_A = \ket{\uparrow}_A$$

If we follow the Copenhagen interpretation, the entire system has actually collapsed to:

$$\ket{\phi'}_{AB} = \ket{\uparrow\uparrow}_{AB}$$

So, assuming they both agreed in advance to measure along the same axis, then Peter knows, with certainty, that Mary measured $\uparrow$ on her particle as well:

$$\ket{\phi'}_B = \ket{\uparrow}_B$$

More importantly, Peter also knows that, had he measured $\downarrow$ instead, Mary would have measured $\downarrow$ instead. This is a counterfactual statement that follows the laws of quantum theory. Since Peter and Mary can be spacelike separated when they perform their measurement, this is a real-world example of a counterfactual dependency in the absence of any causal dependency.

\subsection{Bell's theorem}

One of the earliest original theorems on the non-extensibility of quantum theory comes from \citet{Bell1964} and states that no local \footnote{No signals travel faster than light, i.e., compatibility with special relativity.} hidden-variable theory can reproduce what is predicted by quantum theory. This is because such theories are bounded by the Bell inequalities, and experiments show that nature breaks them. The Bell theorem is based on the EPR experiment.

In such a theory with local realism, the outcomes of any possible measurement would be pre-defined. In our example, Alice and Bob could have measured along the axis

$$(\ket{\nearrow}_X=\frac{\ket{\uparrow}_X+\ket{\downarrow}_X}{\sqrt{2}}, \ket{\searrow}_X=\frac{\ket{\uparrow}_X-\ket{\downarrow}_X}{\sqrt{2}})$$

And obtained either $\nearrow$ or $\searrow$. Such a theory would thus have to define an outcome for all four measurements, which would be pre-existing elements of reality.

It is more accurate today to speak of Bell inequalities, as there is a large number of them \citep{Brunner2014}. A system modelled with local hidden variables is bound by Bell inequalities, but actual experiments can break such inequalities and are thus in contradiction with local hidden variables theories. This discards local realism, and this is known as Bell's theorem.

Formally, Bell inequalities \citep{Colbeck2017} can be defined by modelling the setup with four random variables that take boolean values: A, B, X, Y and considering $P_{XY|AB}$. Another variable $\lambda$ models how the system is prepared. The theory, given by $P_{XY|AB\lambda}$, is left open and its properties are the subject of the discussion.

A causal structure is given: $\lambda \rightarrow A \rightarrow X$ and $\lambda \rightarrow B \rightarrow Y$ and independent free choice is modelled by adding constraints to the underlying probability distribution based on this causal structure, namely: if A is an independent free choice, then whenever another random variable is outside the future light cone of A, A is independent from that other variable. For example $P_{A|\lambda}=P_A$ because $\lambda$ is not in the future light cone of A. 

Based on this assumption, it can be defined what it means for the theory to be locally deterministic (for example, some probabilities must be equal to 0 or 1), and constraints on $P_{XY|AB}$ can be inferred, known as Bell inequalities.

\subsection{The Kochen-Specker theorem}

Later on, the Kochen-Specker theorem \citep{Kochen1967} completed Bell's theorem by replacing locality with non-contextuality \footnote{Elements of reality are independent from the experiments carried out to measure them.}. Non-contex\-tuality is a weaker condition than locality, because it only requires that information on the choices of measurements do not travel faster than light and influence other measurement outcomes. The Kochen-Specker is thus a stronger result. It states that no non-contextual hidden-variable theory can reproduce what is predicted by quantum theory. In other words, it is impossible to assign, before the choice of measurement axis by the experimenting physicist, a value to each observable corresponding to a possible choice of measurement axis, in a way that is consistent, i.e., that it respects functional dependencies between observables. The Kochen-Specker theorem only requires consistency across observables that commute mutually (which is a stronger result than that of Von Neumann).

The independent free choice assumption, coupled with the Kochen-Specker theorem \citep{Kochen1967}, implies that it is mathematically impossible for the outcome of each quantum measurement to be predicted correctly and consistently. If we consider that a physicist may measure, among others, a certain observable, then we cannot predict in general what the result of this measurement will be. Indeed, such an assignment of an element of reality to an observable must be contextual and depend on other, arbitrary measurements that this or other physicists can make on other observables -- at their own, unpredictable, discretion.

\subsection{The Free Will theorem}

This is the essence of the result established by \citet{Conway2006}, known as the Free Will Theorem, under the axioms of FIN, SPIN and TWIN, and without using a probabilistic formalism: if we humans are endowed with independent free choice, then so is the universe. The definition of free will, regarding the physicists freely choosing the measurement axis, is formally stated like so: ``the choice of directions in which to perform spin experiments is not a function
of the information accessible to the experimenters.'' \citet{Conway2006} conclude that ``the free will assumption implies the stronger result,
that no theory, whether it extends quantum mechanics or not, can correctly predict the results of future spin experiments.''

This result was then strengthened by \citet{Conway2009} by weakening the FIN axiom into a MIN axiom, in which only information on the choice of measurement axis cannot travel faster than light, rather than any information. This is similar in spirit to one difference between the Kochen-Specker theorem and Bell's theorem.

\subsection{Maximal informativeness of quantum theory}

\citet{Renner2011} define the strong version of free choice as follows: ``our criterion for A to be a free choice is satisfied whenever anything correlated to A could potentially have been caused by A'', which is another way of stating that A is a free choice whenever it is uncorrelated to anything not its future light cone. Formally, in this definition, A as well as the aforementioned ``anything'' is modelled as a Spacetime Random Variable (SV), which is a random variable with four-dimensional spacetime coordinates. A slightly weaker, non-relativistic version, but still with the same strong idea of independence, only requires for it to be uncorrelated to anything in its past-light cone.

With this strong assumption of independent free choice, which is ``common in physics, but often only made implicitly'' \citep{Renner2011}, coupled with the assumption that quantum theory is correct, they conclude that ``no extension of quantum theory can give more information about the outcomes of future measurements than quantum theory itself.'' In other words, they improve on the past results by showing that, under independent free choice, quantum theory is maximally informative.

In 2018, the Big Bell Test \citep{BigBellTest} involved a large number of people in order to experimentally perform a Bell Test to further confirm the impossibility to extend quantum theory under the existence of free will. For this experiment, it is assumed that the involved experimenters are endowed with free will in the sense that their decisions are ``free variables.''

\subsection{Superdeterminism}

A way out that has been commonly discussed is a superdeterministic theory with no free choice, in which all our choices of measurement axes as well as the corresponding measurement results are predetermined.

For example, Bell made the following statement on a radio interview back in 1985: ``There is a way to escape the inference of superluminal speeds and spooky action at a distance. But it involves absolute determinism in the universe, the complete absence of free will. Suppose the world is super-deterministic, with not just inanimate nature running on behind-the-scenes clockwork, but with our behavior, including our belief that we are free to choose to do one experiment rather than another, absolutely predetermined, including the decision by the experimenter to carry out one set of measurements rather than another, the difficulty disappears. There is no need for a faster-than-light signal to tell particle A what measurement has been carried out on particle B, because the universe, including particle A, already 'knows' what that measurement, and its outcome, will be.''

There is a deterministic, non-local hidden variable interpretation of quantum physics known such as the de Broglie-Bohm theory \citep{Bohm1952}. Its predictive power is identical to other interpretations, as it essentially factors out the randomness into an unknown initial configuration, separating possible worlds in a similar way to Everett's many worlds interpretation. Measurements carried out by experimenters restrict the set of possible initial conditions as we go, but this theory does not explain why the initial conditions of our world are the way they are.

The purpose of this paper is to demonstrate that free choice is not a yes-no matter. There is room in the middle for a weaker free choice assumption: contingent free choice. This leaves open the possibility of theories that are more informative than quantum theory, and in which human agents are still endowed with a non-negligible decisional power.

A particular category of such theories is those that are not only more informative, but also fully deterministic and contextual, while being non-trivial. In such deterministic theories, the decisions of contingently free agents are (epistemically) predictable, albeit not (ontologically) pre-determined, as they have a counterfactual power over the past.

The underlying idea is that, if our universe is indeed absolutely deterministic, then nothing is in the way of us being perfect predictors, i.e., epistemically omniscient -- this is actually the mindset of Newtonian physics, where knowledge of initial conditions allows us to compute everything else. This epistemic omniscience directly interferes with causality and with our utility-maximizing rationality, because our knowledge of the future must still cause, in a special-relativistic sense, that very future (self-fulfilling prophecy). This implies that such a deterministic universe must be the solution to a fixpoint equation. In this fixpoint constraint lies potential for improved predictive power over quantum theory: we can compute information on our world not only through Schrodinger's equation, but through the elimination of the impossible.

This fully deterministic approach is the extreme of a spectrum, but this spectrum also has room in the middle for theories that are more informative than quantum theory while not being fully deterministic. In this paper, we make the case for more informative theories with fully deterministic and transparent theories as an example, because such theories are, paradoxically, simpler to describe and leverage existing game theoretical solution concepts and algorithms, described in Sections \ref{section-extensive-form}, \ref{section-normal-form} and \ref{section-spacetime}. There exist, however, other game theoretical solution concepts such as \citep{Halpern:2013aa} based on translucency, that drop independent free choice for contingent free choice, while retaining uncertainty and probabilities in their framework. Such theories may be useful to find more informative, but non deterministic extension theories that may be more conducive to actual experimental setups.

Contingent free choice is not the complete absence of free will, because we could still have acted otherwise and bear responsibility on what our decisions entail. But contingent free choice is also not independent free choice: it models that we act consistently and rationally, taking our environment into account.


\section{Perfect prediction and rationality in all possible worlds with timelike separation}
\label{section-extensive-form}
\subsection{Initial conjecture}

The weakening of the independent free choice assumption to the contingent free choice assumption was explored in depth in the field of game theory and rational choice. \citet{Lewis1979}, \citet{Pettit1988} and \citet{Dupuy1992} pointed out an analogy between Newcomb's problem and the Prisoner's dilemma.

\citet{Dupuy2000} then suggested a new approach to rational choice theory based on the notion that the prediction of a rational agent's decision is correct in all possible worlds, and that all agents are rational in all possible worlds. Being rational, in this context, means that given the choice between several outcomes, the agent will select the outcome that they prefer the most. Being rational in all possible worlds means that, even if something (anything) had been different, the agents would still have been rational: this is a counterfactual statement as we introduced them in Section \ref{section-counterfactuals}. Likewise, a correct prediction in all possible worlds means that, even if something (anything) had been different, the agents would still have correctly predicted each other.

Dupuy gave a few examples on a few simple games known as take-or-leave as well as centipede games, and conjectured that under these assumptions, the underlying solution concept, which he called Projected Equilibrium, always exists, is unique, and is always Pareto-optimal. Pareto-optimality is the desirable feature in economics that no other possible outcome of the game gives a better payoff to all players, i.e., that the equilibrium is never suboptimal.

\subsection{Perfect prediction equilibrium}

The solution concept was formally defined for all games in extensive form with perfect information and in general position (no ties) in 2004 by \citet{Fourny2018} as the Perfect Prediction Equilibrium and the three conjectures were proven. The key to the reasoning is the use of a forward induction mechanism -- in contrast to the Nashian backward induction -- that eliminates outcomes one by one until the last one remains. An outcome is eliminated -- we also say: preempted -- if its own prediction causes a deviation to a different, incompatible subtree. In other words: all outcomes subject to a Grandfather paradox are eliminated. The Perfect Prediction Equilibrium is the only outcome that is immune to its prediction, in the sense that, knowing it in advance, the players play towards this very outcome.

From a logical perspective, the Perfect Prediction Equilibrium is the solution of a fixpoint equation going backward and forward in time, namely, that the outcome must be caused by its prediction, the latter being counterfactually dependent on it.

The main idea underlying perfect prediction is that, if agents can indeed predict each other in all possible worlds, then this induces consistency constraints over what the actual world can look like because of the way counterfactual dependencies interfere with causal dependencies. The actual world must indeed be consistent with both a correct prediction and a consistent timeline.

\begin{figure}
\includegraphics[width=8.6cm]{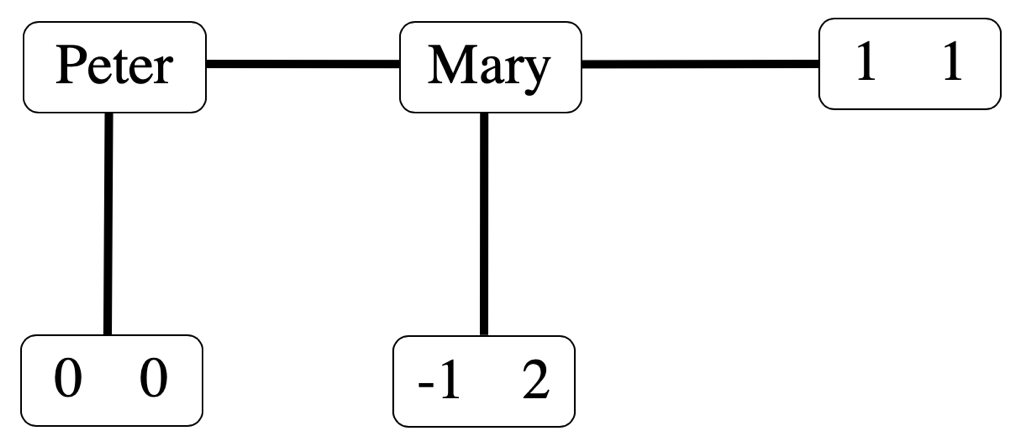}
\caption{A game in extensive form. The Subgame Perfect Equilibrium, which is a Nash Equilibrium, is obtained with a backwards induction and is (0,0). In this Nashian framework, Mary's decision to pick (-1,2) if she were to play, even though if does not actually happen, is an ``element of reality'' on which Peter's decision to pick (0, 0) is based. In this respect, Nashian game theory is non-contextual because it assigns a decision to every node, even outside the equilibrium path.}
\label{figure1}
\end{figure}
Figure \ref{figure1} shows a simple game in extensive form, called the promise game. Peter is a baker who may or may not give a loaf of bread to Mary, who then may or may not pay. The numbers correspond, on the left, to what Peter gets, and on the right, to what Mary gets.

Under the Nash setting with independent free choice, the asynchronous exchange does not happen: if Mary were given the bread, it would not be in her interest to pay. Knowing this, Peter does not give her anything.

It should be noted that the Nash reasoning, and backward induction in general, is explicitly non-contextual: indeed, even though Mary does not actually get to decide anything in the actual world, her strategy is ``assigned'' as not paying, and Peter's reasoning is based on this hypothesis. This is similar in quantum theory to the assignment of measurement outcomes to each possible choice of the measurement axis in local realism, which leads to contradictions\footnote{Bell inequalities in the local realism case, backward induction paradox in the Nash case}.

\begin{figure}
\includegraphics[width=8.6cm]{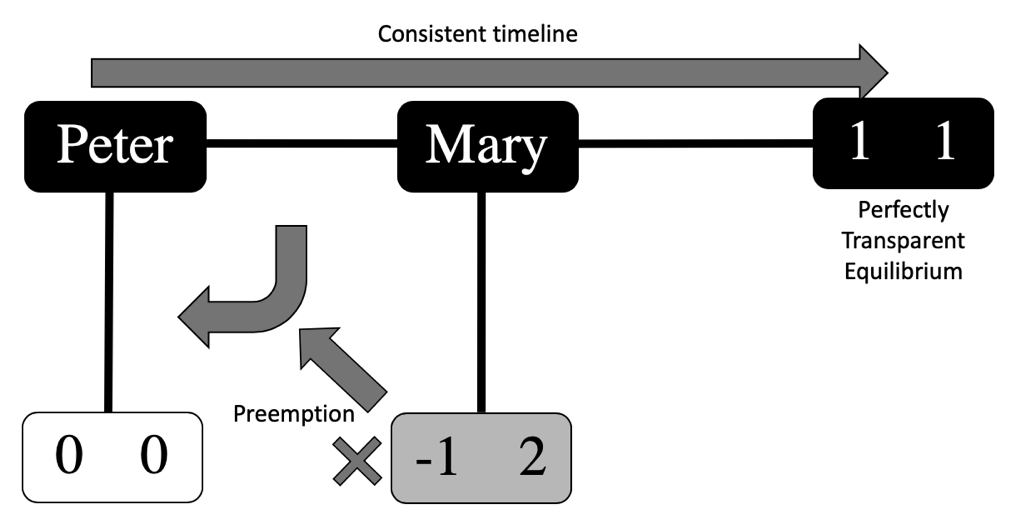}
\caption{The Perfect Prediction Equilibrium. (-1,2) is preempted, and (1,1) is the solution of this game under rationality in all possible worlds and perfect prediction. This equilibrium is contextual because of its counterfactual structure: it only assigns decisions to nodes that are actually reached, even for bigger games, and the decisions \emph{would have been different} if other nodes had been reached.}
\label{figure2}
\end{figure}

Figure \ref{figure2} shows the resolution of this game under contingent free choice and assuming players are rational in all possible worlds as well as perfect predictors. Outcome (-1, 2) cannot be known as the solution in the game: if that were the case, Peter would have known, and would have deviated to (0,0) because $0 > -1$. This is a reductio ad absurdum reasoning that eliminates the path leading to (-1,2) as an inconsistent timeline, not immune against its own anticipation. Having eliminated (-1, 2), Peter finds it in his best interest to hand over the loaf of bread to Mary because he knows (1,1) is the only logically consistent outcome that can then happen. Mary then picks (1,1), because she knows that, if she picked (-1, 2) instead, Peter would have known it, and she would not have been in a position to make a decision at all \footnote{For the reader for whom this is too strong a statement: Mary's behavior can also be seen as a kind of categorial imperative willingly followed by an agent, in the sense envisioned by Immanuel Kant. It is enough that Mary believes that this counterfactual implication holds, in order to behave suchly.}

The Perfect Prediction Equilibrium is contextual in the sense that only decisions that are on the equilibrium path are assigned, all others (non reached nodes) being undefined. On this small example, all decision nodes are on the path to (1,1), but on more complex games, this is not the case. This is close to the spirit of quantum theory that the position of a particle may be measured and defined, while its speed is undefined\footnote{More generally, it is not possible to jointly measure two observables that do not commute.}: the decision of an agent is only defined if they actually get to make this decision.

\section{Perfect prediction and rationality in all possible worlds with spacelike separation}
\label{section-normal-form}

The twin solution concept for games in normal form, which can be played by spacelike-separated players in separate rooms, was formalized in 2017 \citep{Fourny2017} as the Perfectly Transparent Equilibrium. It is based on the iterated elimination of non-individually rational outcomes. It follows the same logics as its extensive-form counterpart. However, even though it is unique and Pareto-optimal as well, it does not always exist.

An epistemic characterization was given \citep{Fourny2018b} by formalizing the concepts of rationality in all possible worlds (formally called necessary rationality) and perfect prediction (formally called necessary knowledge of strategies, knowledge of strategies in all possible worlds) into Kripke semantics. The latter can also be referred to as a form of epistemic omniscience, in the sense that the agents know all the events that happen in their world.

\begin{figure}
\includegraphics[width=8.6cm]{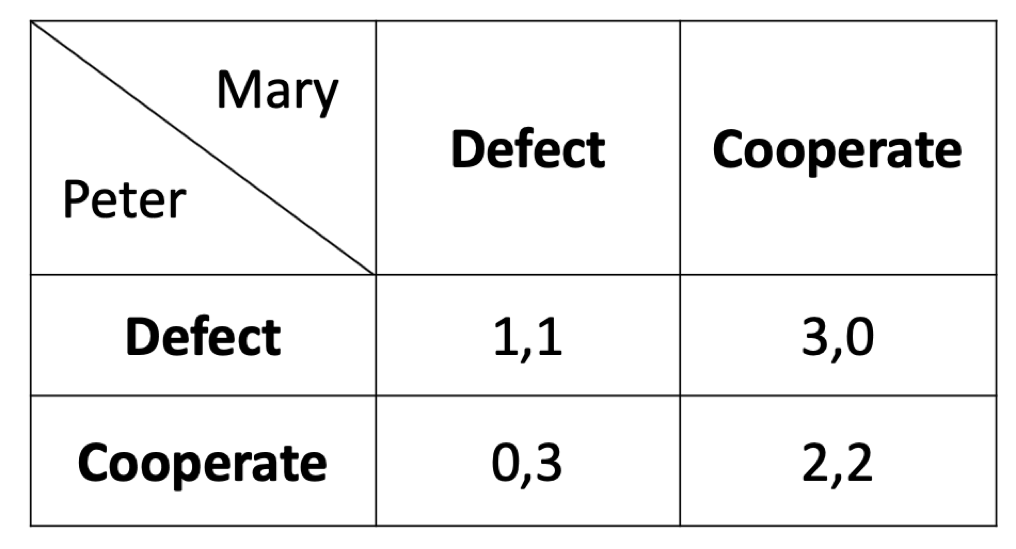}
\caption{The prisoner's dilemma, a game in normal form played by spacelike-separated agents. The Nash equilibrium is (1,1) because under independent free choice, defecting is a best response to the other player's fix decision, no matter what it is.}
\label{figure3}
\end{figure}

Figure \ref{figure3} gives a simple example with the prisoner's dilemma, involving two agents that can either cooperate or defect. The difference with the previous games is that Peter and Mary are in separate rooms and cannot communicate: they are space-like separated.

Cooperating when the other player defect is punished by a low payoff, inducing a dilemma. The Nash solution is (1,1), because if players have independent free choice, payoffs are compared across rows or columns, and each player has no interest in deviating.

\begin{figure}
\includegraphics[width=8.6cm]{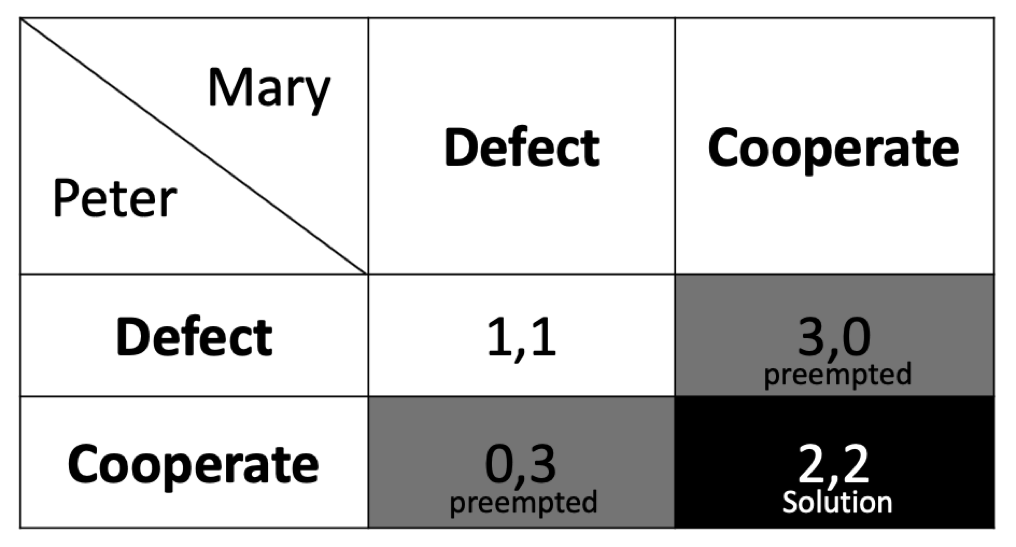}
\caption{The Perfectly Transparent Equilibrium. (0,3) is preempted as well as (3,0). Knowing this, (2,2) is then immune to its own anticipation by both agents and is the solution.}
\label{figure4}
\end{figure}

Figure \ref{figure4} shows the resolution of this game under contingent free choice and assuming players are rational in all possible worlds as well as perfect predictors. First, (0,3) and (3,0) are shown not to be logically consistent under these assumptions, since a player who would anticipate to get 0 would deviate to their other strategy to guarantee a payoff of 1. Having eliminated these two outcomes, the comparison is done across the diagonal, leading to (2,2) as the solution immune to its own anticipation.

Note that \citet{Hofstadter1983} made an argument leading to the same solution for symmetric games, called superrationality. The Perfectly Transparent Equilibrium generalizes this to any games in normal form, even non symmetric.

There are numerous other papers published in Game Theory with similar, weakening approaches to free choice and that do not assume full determinism: for example, \citet{Halpern:2013aa} researched what happens when agents are translucent, which means that in contrast to perfect prediction, some information leaks but not all of it. \citet{Shiffrin2009} has another non-Nashian approach to games in extensive forms, that one could also call translucent. These other approaches, in contrast to the PPE and PTE, consider other possible worlds not to be impossible possible worlds. However, \citet{Shiffrin2009}'s approach, in extensive form, is non-contextual and follows the Nashian approach in that it assigns a strategy to every node, even if unreached. An interesting avenue of research would be to find solution concepts that are not fully deterministic, but that are nevertheless contextual. \citet{Halpern:2013aa}'s work on the normal form goes into this direction in that they leave decisions partly undefined.

\section{Perfect prediction and rationality in all possible worlds in a generic spacetime setting}
\label{section-spacetime}

\subsection{Perfect information and Minkowski spacetime}

\citet{Fourny2019} generalized the equilibrium to games in extensive form with imperfect information, which are a superset of both games in normal form and games in extensive form with perfect information. It was shown that this broader class of games is adequate to model any decisional setup across Minkowski spacetime\footnote{There are thus no closed timelike curves}, where the decisions of the agents may be, at will, spacelike or timelike separated.

When all agent decisions are spacelike separated, this comes down to a game in normal form in which agents decide in separate rooms, for which the equilibrium is defined in Section \ref{section-normal-form}. When all agent decisions are timelike separated, and thus a decision only takes place if decisions in the past were made in a certain way, this comes down to a game in extensive form with perfect information, for which the equilibrium is defined in Section \ref{section-extensive-form}.

\begin{figure}
\includegraphics[width=0.6\textwidth]{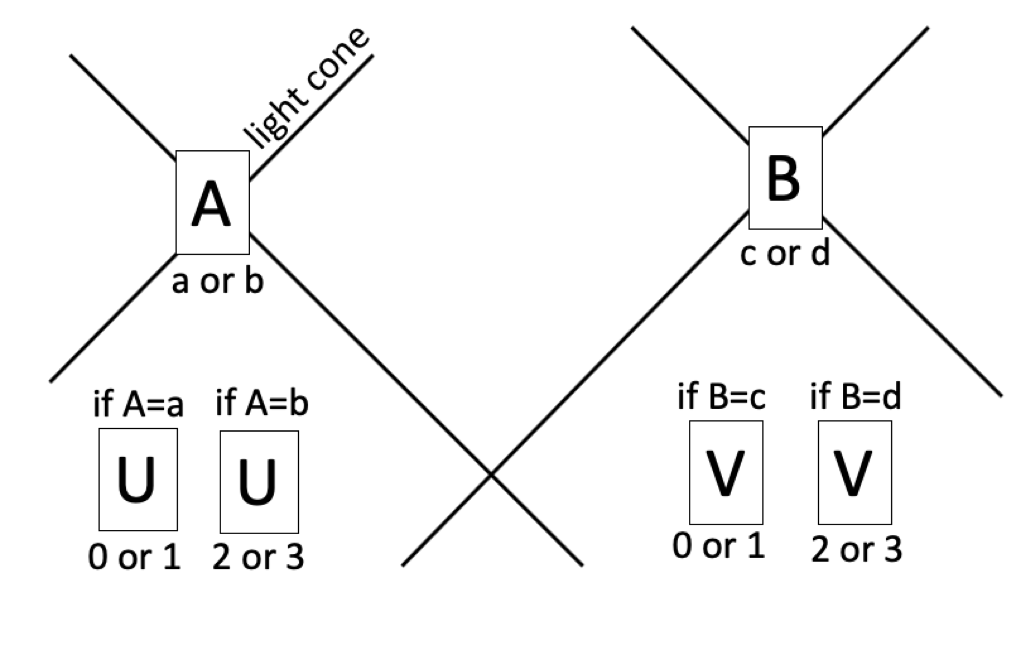}
\caption{Four agents making decisions in spacetime: A, B, U and V. The light cones are indicated and signals cannot travel faster than light.}
\label{figure5}
\end{figure}

Figure \ref{figure5} shows an example where agents are organized in a mixture of spacelike-separated and timelike-separated locations with decision points. When two decision points are timelike-separated, the later decision only actually takes place if the former decision was taken in a certain way. In this example, A can pick a or b and B can pick c or d, in a spacelike-separated manner. U and B, as well as V and A are spacelike-separated, meaning that U's decision points are only affected by A, and V's decision points are only affected by B. If A picked a, then U gets to pick between 0 and 1. If A picked b, then U gets to pick between 2 and 3. If B picked c, then V gets to pick between 0 and 1. If B picked d, then V gets to pick between 2 and 3. 

The keen reader may recognize here that this above setup is actually that of the EPR experiment, where two physicists pick measurement axes, and the universe picks measurement outcomes. For now though, we consider that these are just four agents making decisions.

There are 16 different possible outcomes with this setup. For example, one of them is that A picks a, B picks d, U picks 1 and V picks 3. It is assumed that the agents have a total preference order on this 16 different outcomes. In this example, we also assume that U and V have the same preference order\footnote{Making U and V two different agents makes it clear that no signals are sent across spacelike-separated locations, where the measurement outcomes are determined.}.

\begin{figure}
\includegraphics[width=\textwidth]{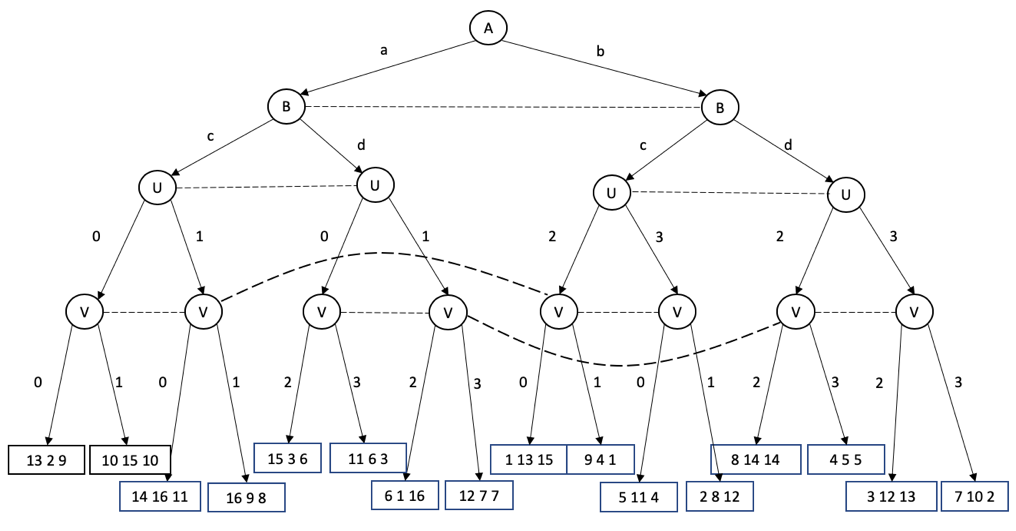}
\caption{A game in extensive form and with imperfect information involving four players: A, B, U and V. The dotted lines represent the 6 information sets: A deciding between and b, B deciding between c and d, U deciding between 0 and 1, U deciding between 2 and 3, V deciding between 0 and 1, V deciding between 2 and 3.}
\label{figure6}
\end{figure}

Figure \ref{figure6} shows how this setup is mapped to a game in extensive form with imperfect information, following the algorithm provided by \citet{Fourny2019}. Because A and B are spacelike separated, the two nodes at which B makes a decision (left if A picks a, right if A picks b) are part of the same information set, which is indicated with the dotted line. Being in the same information set meaning that it is not known at which node one is within this information set. In other words, B's decision must be made with no direct knowledge about A's decision.

Likewise, U only gets to pick between 0 and 1 if A picked a. U only gets to pick between 2 and 3 if A picked b. V only gets to pick between 0 and 1 if B picked c. V only gets to pick between 2 and 3 if B picked d.

The 16 leaves of the game correspond to the 16 possible outcomes. The numbers are payoffs with an ordinal meaning and encode the preference functions: The first number is A's payoffs, the second number is B's payoff, the third numbers is both U's and V's payoff assuming their preference relation is the same.

\subsection{Computation of Nash Equilibria}

\begin{figure}
\includegraphics[width=\textwidth]{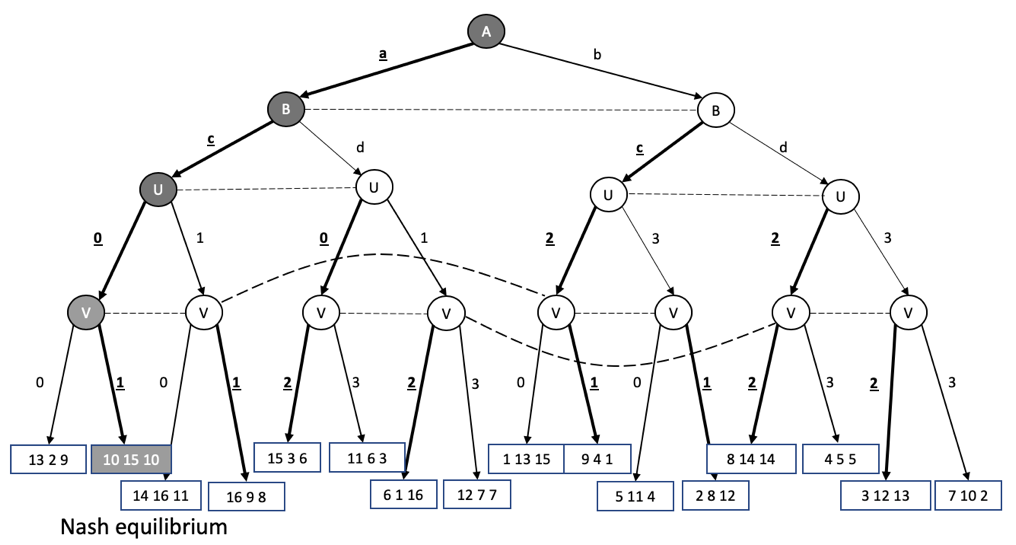}
\caption{A Nash equilibrium for the example game. Each information set is assigned a decision, leading to the outcome (10, 0, 10). No agent has any interest unilaterally deviating from their strategy, as this would diminish they payoff. In this equilibrium, the six decisions are as follows: A picks a, B picks c, U picks 0 and 2, and V picks 1 and 2. Note that U and V picking 2 is defined even though not on the path the equilibrium outcome. The Nash paradigm is thus non-contextual.}
\label{figure7}
\end{figure}

Figure \ref{figure7} shows how the classical resolution of the game would work according to Nashian game theory. A Nash equilibrium is an assignment of a choice to each decision point \footnote{It is important to note that this assignment must be the same for nodes connected with a dotted line, also called an information set. This is the meaning of imperfect information, due to some spacelike-separation in the decisional setup.}

In general and including in this example, there may be an arbitrary number of Nash equillibria, also possibly zero for some games.

Outcome (10, 15, 10) is, for example, a Nash equillibrium because:

\begin{itemize}
\item if A had picked b instead, all other decisions unchanged, she would have gotten 9 instead of 10.
\item if B had picked d instead, all other decisions unchanged, he would have gotten 3 instead of 15.
\item if U had picked 1 instead (and any of 2 or 3), all other decisions unchanged, he would have gotten 8 instead of 10.
\item if V had picked 0 instead (and any of 2 or 3), all other decisions unchanged, he would have gotten 9 instead of 10.
\end{itemize}

No agent thus has any interest in unilaterally deviating from their chosen strategy. Note that the assignments are non-contextual: V picks 2 over 3 and U picks 2 over 3, although these decisions are not actually on the timeline If an agent unilaterally deviates, all other assignments would have been the same by definition of the Nash equilibrium and independent free choice, which is the essence of non-contextuality.

\subsection{Computation of the Perfectly Transparent Equilibrium}

\begin{figure}
\includegraphics[width=\textwidth]{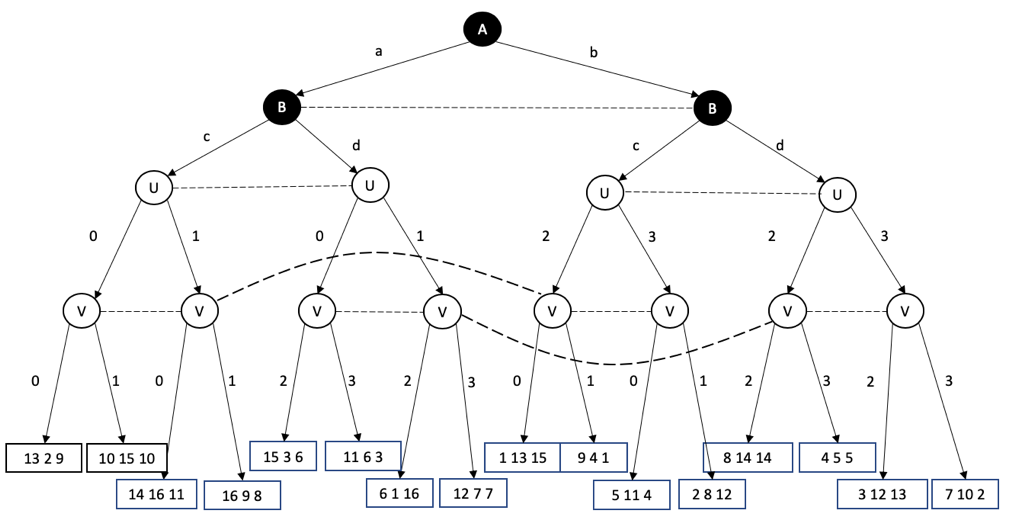}
\caption{The initialization of the computation of the Perfectly Transparent Equilibrium. No matter what, A and B get to make a decision, which is why they are marked in black.}
\label{figure8}
\end{figure}

In order to compute, on the other hand, the Perfectly Transparent Equilibrium, we start by observing that A and B get to make their decision no matter what. However, we do not know yet for U and V, as the decision they get to make depends directly on what A and B have decided before. On Figure \ref{figure8}, we have marked A's and B's decision nodes in black.

\begin{figure}
\includegraphics[width=\textwidth]{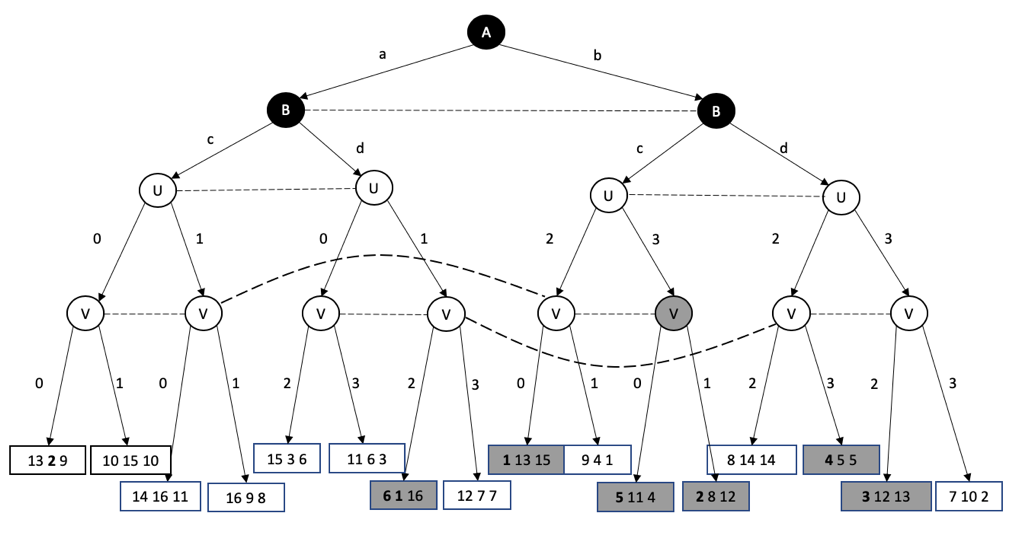}
\caption{The first round of elimination in the computation of the Perfectly Transparent Equilibrium. The six outcomes marked in gray correspond to inconsistent timeline: if it were the solution, either A or B would had deviated, which is a form of Grandfather's paradox. }
\label{figure9}
\end{figure}

The next step is to observe that, if A picked a, the worst that could happen no matter what anybody else is doing is that she gets 6. It follows directly that (1, 13, 15) cannot be the final outcome under rationality in all possible worlds and perfect prediction: if (1, 13, 15) is the actual solution and A knows it in advance, then A's rational choice is to pick a because $6>1$. However, A picking a cannot cause \footnote{Reminder: causality is based on light cone semantics.} (1, 13, 15) as it can only be caused by A picking b. This is a form of Grandfather's paradox. (1, 13, 15) is thus eliminated because it is impossible under the assumptions. The same goes for five other outcomes, marked in gray on Figure \ref{figure9}. A sixth outcome (6, 1, 16) is also eliminated at the same step, because the other agent, B, would have deviated to c (securing a guaranteed minimum of 2), if it had been the solution.

\begin{figure}
\includegraphics[width=\textwidth]{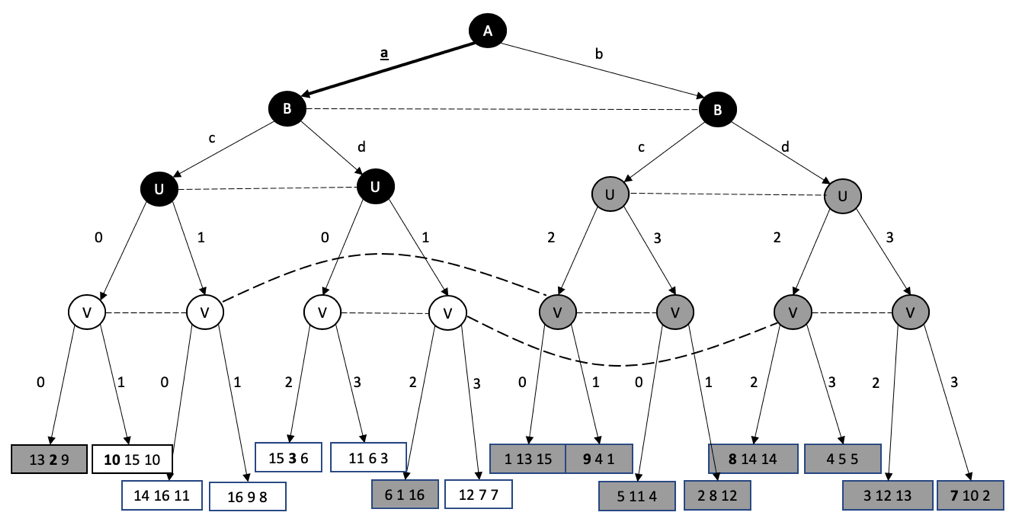}
\caption{Another round of elimination, based on the first one. Three more outcomes cannot be caused by their anticipation by rational players A and B, and thus are eliminated. After this second round, we know that A's decision is a, and we know that U gets to make a decision between 0 and 1 no matter what.}
\label{figure10}
\end{figure}

\begin{figure}
\includegraphics[width=\textwidth]{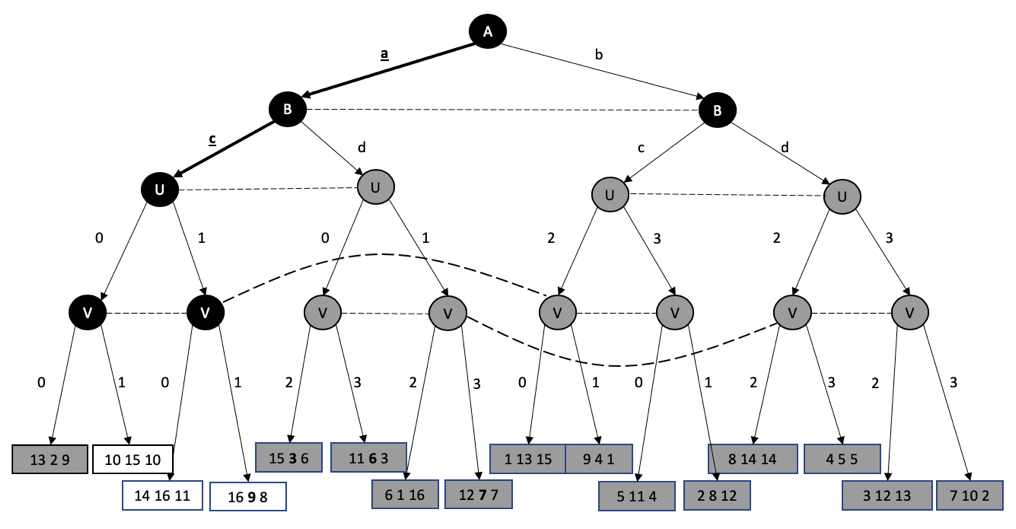}
\caption{Another round of elimination, based on the second one. Three more outcomes cannot be caused by their anticipation by rational players, B and U and thus are eliminated.  We know after this round that B's decision is c, and that V gets to choose between 0 and 1.}
\label{figure11}
\end{figure}

\begin{figure}
\includegraphics[width=\textwidth]{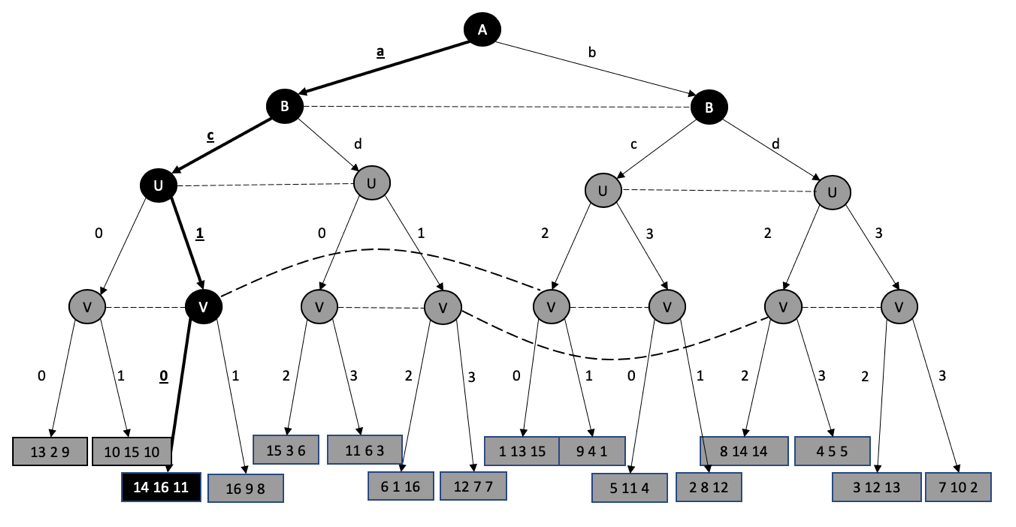}
\caption{A last round of elimination. At that point, only one outcome survives. It is the Perfectly Transparent Equilibrium. It is contextual and decisions are only defined on its path.}
\label{figure12}
\end{figure}

Figures \ref{figure10}, \ref{figure11} and \ref{figure12} show the next steps, as more outcomes are eliminated and as the forward induction progresses down the tree to U and V, and until at most a single outcome remains.

A quantum physicist will thus see here that the assignments, under contingent free choice, are contextual, whereas the Nash resolution under independent free choice is non-contextual. Indeed, unlike a Nash resolution, a strategy does not assign a decision to all information sets, but only to those on the equillibrium: (14, 16, 11) is the Perfectly Transparent Equilibrium. Anticipating this, A picks a, B picks c, U picks 1 and V picks 0, which leads to precisely this outcome (fixpoint). U's decision between 2 and 3 and V's decision between 2 and 3 remain \emph{undefined} and are irrelevant to the reasoning. These two undefined decisions are no elements of reality: only what is actually happening on the timeline is an element of reality.

\clearpage

\section{A fixpoint-based theory of physics with more predictive power}
 \label{section-theory}
 
The purpose of this paper is to show that contingent free choice provides a way out from the impossibility to extend quantum theory to a theory with improved predictive power \citep{Renner2011}.

In this section, we provide an actual (parameterized) candidate theory based on the Perfectly Transparent Equilibrium obtained in a game of imperfect information played between, on the one side, agents endowed with contingent free choice and, on the other side, the universe.

We emphasize that this theory is one possible theory, but the weakening of independent free choice into contingent free choice unlocks a wider class of theories with potentially improved predictive power.

\subsection{Link between game theory and quantum theory}

There are a number of similarities and common concepts between game theory and quantum theory.

The first commonality is that the notion of possible worlds is present both in Kripke structures, with a set of possible worlds equipped with an accessibility relation, and in quantum theory with the possible measurements that can be carried out, the possible measurement outcomes, and thus the possible states that the system can assume in the Hilbert space. In the many-worlds interpretation \citep{Everett1973}, possible worlds are explicitly part of the model. In the De Broglie-Bohm pilot-wave theory, possible worlds correspond to trajectories, indexed on the space of possible initial conditions.

Secondly, based on these possible worlds, the notion of counterfactuals is at the core of quantum theory: the outcome of a measurement was 1, but it could have been 0. The cat is alive, but it could have been dead. Quantum theory deeply embeds the notion of unrealized possibles in its mathematical framework. Whether these unrealized possibles are real or not is the subject of intense debates between supporters of the Copenhagen interpretation and of the Many-worlds interpretation \citep{Everett1973}. Quantum theory also has, as its core, counterfactual dependencies in the way envisioned by \citet{Lewis1973}, which directly appear in statements such as ``If Alice had measured $\downarrow$, then Bob would have measured $\downarrow$.'' (see also \citet{Fuchs2010}).

Thirdly, the Nash resolution of a game corresponds to the assignment of a decision to each decision point and means that the decision of an agent is predefined for each possible decisions the other agents can make in the past.. This is the same idea as the measurement outcomes of a system being pre-defined for each possible measurement that can be carried out in local hidden variable theories.

Giving up local realism is often described as quantum theory being contextual: measurement outcomes are only assigned consistently to measurements that are actually carried out, and they depend on which measurements are carried out: a given measurement would have given a different outcome if the \emph{other} measurements had been done differently.

As we pointed out, the Perfectly Transparent Equilibrium approach, based on contingent free choice, is contextual: the only decision outcomes that are defined are those on the equilibrium path, i.e., those that are actually made. Any other decisions are undefined and not part of the reasoning. The Perfectly Transparent Equilibrium framework thus provides an alternate theory for experiment such as the EPR experiment, parameterized on utilities \footnote{Speaking of a class of theories may be more accurate, as the theory can be instantiated with various models of utilities, which is future work.}. This theory is fully deterministic, contextual \footnote{It thus does not fulfil local realism in the sense of \citet{Bell1964}.}, nonlocal \footnote{In the sense that counterfactual dependencies are not constrained by the speed of light; however, it is in some sense ``local,'' in that no information can be sent faster than the speed of light.}, and non-trivial, and corresponds to the mindset of one-boxers in the Newcomb problem. We believe there may also be theories ``inbetween" quantum theory and such a fully deterministic theory, that is, that are more informative but not fully informative, but leave it for further research.

\subsection{The Perfectly Transparent Equilibrium used as an extension theory of quantum theory}

The EPR experiment can be modelled as a game in extensive form with imperfect information, where A, B, X and Y are choices made by agents and the universe. This setup was already shown on Figure \ref{figure5}, and this game has the structure of the one shown on Figure \ref{figure6} -- with the utilities between an external parameter.

More generally, any experimental or theoretical setup in which physicists located at certain positions in spacetime pick measurement axes, carry out measurement and find specific outcomes, can be recast into a game in extensive form with imperfect information. The algorithm to do so was fully formalized by \citet{Fourny2019} and we illustrate this algorithm with the EPR game in this paper. The game is built given the causal structure of the experimental setup, based on spacelike and timelike separation, as well as the contingency constraints (measurement X is only carried out if this and that happened before). Note that this, however, does not cover hypothetical and theoretical setups such Winger's friend \citep{Winger1961} or the Frauchiger-Renner experiment \citep{Frauchiger2018}, in which human beings may be part of non-classical states.

In other words, if we assume contingent free choice rather than independent free choice, then the story this theory tells us about quantum experiments is this:

\begin{itemize}
\item The choice of the measurement axis is known in advance (correct prediction)
\item Values are only assigned to observables for those known choices of measurement axes, and are an element of reality. Other observables are not assigned any outcome and are undefined.
\item The physicist could have picked a different axis (contingent free choice)
\item The choice of measurement axis would also have been known if the physicist had picked a different axis (which implies dropping independent free choice).
\end{itemize}

The utility of the game are parameters. Unlike the De Broglie-Bohm theory, these are not arbitrary initial conditions, but correspond to the utilities of rational agents, the physicists, as well as any quantity that the universe optimizes; many theories involve the minimization of a quantity: shortest path taken by light, quantum potential, etc. The definition of the utility of the universe is the natural next step in future work. We expect the utility of the universe to be governed by the Schrodinger equation or something akin to the quantum potential in the De-Broglie-Bohm theory.

With the additional assumptions of perfect prediction and rationality in all possible worlds, the simple theory presented in this paper excludes possible worlds in which a different choice of (and correct prediction of) measurement axis would counterfactually lead to an inconsistent world (Grandfather's paradox...). In Kripke semantics, this is known as an impossible possible world \citep{Kripke1963}\citep{Rantala1982}. Eliminating inconsistent worlds in a way similar to game theory (PPE, PTE) can thus narrow down which actual worlds are allowed by the laws of physics, possibly only one. The theoretical ability to be able to narrow down possible worlds to just one is precisely closing the feedback look, as envisioned by \citet{Dupuy2000}: if we can do so, then we can compute in theory the choice the physicist will make, and even further, we would also have computed the choice the physicist would have made, had it been different.

The Perfectly Transparent Equilibrium theory is bootstrapped by first assuming that there exists a deterministic theory that tells us which world is the actual world, concluding that this world must fulfil constraints such as perfect prediction (because the theory is deterministic) and rationality in all possible worlds, deriving this theory, and seeing that this theory predicts that this world is at most unique, validating the consistency of the perfect prediction assumption.

Another way to view this theory is that it is a contextual, deterministic, nonlocal hidden variable theory, and the hidden variables can be calculated as solutions of the fixpoint equation modelled by the Perfectly Transparent Equilibrium. This theory, or class of theories if we consider that it is parameterized by utilities, is essentially an augmented Everettian many-world theory, in which the actual world (or part of it) is not governed by probabilities, but is instead necessary and can be computed, based on the additional constraints entailed by the postulated predictability of the choice of measurement axis.

\section{Next avenues of research}

\subsection{Reproducing the statistical results predicted by quantum theory}

An extension theory of quantum theory must be able to reproduce results predicted by quantum theory, in particular, the statistical distribution of measurement outcomes. It must account for the statistical distributions we actually observe in experimental setups (probability of measurement outcomes conditioned on chosen measurement axes, see \citep{Colbeck2017}). Since the theory is deterministic, it means that these statistical distributions must be due to our imperfect knowledge of the world -- in the De-Broglie Bohm theory, it is similar to not knowing the initial conditions.

The deterministic theory presented in the Section \ref{section-theory} is parameterized by utilities. A statistical distribution of measurement outcomes could be explained by incomplete knowledge on these utilities: if rather than a single game, we consider random games with the same structure but different utilities, then a statistical distribution of measurement outcomes can be theoretically inferred.

Consequently, further work includes considering games where the utilities are drawn at random (for example, modelling uncertainty or incomplete knowledge), looking at the distributions of final outcomes, comparing them with the predictions made by quantum theory, investigating how the distribution of randomly drawn games maps to the distribution of outcomes, and investigating which distributions of outcomes can be obtained in general.

\subsection{Alternate extension theories based on contingent free choice}

The theory presented in Section \ref{section-theory} is meant as an example to illustrate how weakening the free choice axiom unlocks new classes of theories with increased predictive power,

In the future, further theories also assuming contingent free choice may be built with weaker assumptions than perfect prediction and rationality in all possible worlds, as these latter two assumptions require contingent free choice to make any sense, but the converse is not true: contingent free choice does not require perfect prediction and rationality in all possible worlds (translucent settings \`a la \citet{Halpern:2013aa}). As it turns out, assuming the agents are endowed with perfect prediction \footnote{As Dupuy pointed out, it actually suffices that the agents \emph{believe} that everybody else is a perfect predictor. This assumption is thus less strong than it seems.} and rationality in all possible worlds makes the theory \emph{simpler}.

\subsection{Experimental setups}

Independent free will is defined formally, and thus it is the case that nature either obeys independent free will or does not. An instantiation of the Perfectly Transparent Equilibrium theory with given utilities, based on contingent free will, has a stronger predictive power than quantum theory in its current state, because it predicts which measurement axes are chosen and which outcomes each measurement yields. It is thus falsifiable. It may be within our technological reach to design an experimental setup that can confirm or deny instantiations of this theory with utility models, independently of matters of taste or of personal opinions on the debated topic of free choice. Such experiments would be accessible to us if either the hypothetical, global fixpoint equation can be solved partially in certain closed setups, or if we can build alternate, non-fully-deterministic theories that drop independent free choice while not being fully informative, lowering the bar to building an experiment.

The most obvious starting point for designing experimental setups is to let the universe play against itself, i.e., re-use measurement outcomes as choices of measurement axes.

\section{Conclusion}

We started with the premise, demonstrated in literature, that the current limitation in extending quantum theory into a more informative theory is due to the assumption of independent free choice. We showed that this assumption can be weakened to contingent free choice, in which the decisions of agents may be counterfactually or statistically dependent on events not in their future light cone. We illustrated, with a candidate deterministic theory, the feasibility and consistency of this approach in the ideal context of a perfect knowledge of the world.

We submitted in particular that our decisions on what we want to measure, and the universe's choice of the outcome that we observe as a result, may be the one and same physical, contingency phenomenon; the theory presented in this paper does not make any distinction.

\section{Acknowledgements}

Philosophical credits go to Jean-Pierre Dupuy, who laid down the philosophical foundations of projected time, underlying the Perfect Prediction Equilibrium and the Perfectly Transparent Equilibrium. Jean-Pierre pointed out at multiple occasions that there is a strong link between Newcomb's problem and Schr\"odinger's cat, and that this should be investigated. I am also thankful to St\'ephane Reiche, with whom I collaborated on building the algorithmic framework behind the Perfect Prediction Equilibrium.

The general idea presented in this paper was mentioned for the first time in a talk on the Perfect Prediction Equilibrium that I gave back in 2009 at a lunch seminar hosted by the ETH Zurich institute for theoretical physics. I also had exciting and motivating discussions at the Solstice of Foundations, hosted at ETH Zurich in 2017 for the 50th anniversary of the Kochen-Specker theorem, as well as at the ASIC 2017 and ASIC 2019 conferences organized by Rich Shiffrin.

I am thankful to Renato Renner for various exchanges of emails as well as a few offline discussions on the matter, and to Roger Colbeck and other members of the group for pointing me to related papers in physics and getting me started. I am also indebted to Marcello Ienca for numerous conversations on the neuroscience aspects of free will. Marcello gave me pointers to many relevant papers on this topic. I had further fruitful conversations with Bernard Walliser, Rich Shiffrin, Joe Halpern, Bob French, Matt Jones, Jerome Busemeyer, Harald Atmanspacher.

It has to be said that some of my colleagues named in the above paragraphs are convinced two-boxers and proponents of independent free choice; their open mind honors them and makes the discussions even more exciting.

\bibliographystyle{unsrtnat}

\end{document}